\documentclass[aps,pra,twocolumn,nopacs,floatfix]{revtex4}
\usepackage{epsfig}
\usepackage{graphicx}
\usepackage{dcolumn}
\usepackage{amsthm,amsmath}
\usepackage{comment}
\usepackage[colorlinks]{hyperref}
\usepackage{xcolor}
\usepackage{longtable}
\usepackage{amsmath}

\usepackage{morefloats}
\usepackage{ulem}
\usepackage{verbatim}

\begin{document}

\title{\bf{Two- and three-body dispersion coefficients for interaction of Cu and Ag atoms with {Group} I, II, and XII elements}}

\author{$^{1,2}$Harpreet Kaur}
\email{kaurhpd@gmail.com}
\author{$^3$Jyoti}
\author{$^4$Neelam Shukla}
\author{$^{1,5}$Bindiya Arora}
\email{bindiya.phy@gndu.ac.in}

\affiliation{$^1$Department of Physics, Guru Nanak Dev University, Amritsar, Punjab 143005, India}
\affiliation{$^2$Department of Physics, Khalsa College, Amritsar, Punjab 143002, India}
\affiliation{$^3$Institute of Quantum Electronics, School of Electronics, Peking University, Beijing 100871, P. R. China}
\affiliation{$^4$Department of Physics, University of Nebraska at Kearney, NE-68849, USA}
\affiliation{$^5$Perimeter Institute for Theoretical Physics, Waterloo, Ontario N2L 2Y5, Canada}

\begin{abstract}
The mounting interest in conducting thorough analyses and studies of long-range interactions stems from their wide-ranging applications in cold atomic physics, making it a compelling area for research. In this work, we have evaluated long range van der Waals dispersion (vdW) interactions of Cu and Ag atoms with atoms of group I (Li, Na, K, Rb, Cs, and Fr), II (Be, Mg, Ca, Sr, and {Ba}), XII (Zn, Cd, and Hg) {as well as} singly charged ions of group II (Be$^+$, Mg$^+$, Ca$^+$, Sr$^+$, and {Ba$^+$}) and XII (Zn$^+$, Cd$^+$, and Hg$^+$) by calculating $C_6$(two-body) and $C_9$ (three-body) vdW dispersion coefficients. In order to obtain these $C_6$ and $C_9$ coefficients, we have evaluated the dynamic dipole polarizability of the considered atoms using appropriate relativistic methods and the sum-over-states approach.
To ascertain the accuracy of our results, we have compared the evaluated static dipole polarizabilities of Cu and Ag atoms and their oscillator strengths for dominant transitions {with} available literature. The calculated values of $C_6$ dispersion coefficients have also been compared with the previously reported results.
\end{abstract}
\maketitle

\section{Introduction}\label{sec1}
In the past few decades, considerable attention has been paid to long-range interactions among various atomic species due to their ubiquitous nature, which plays a critical role in the realization of different aspects of cold atom physics \cite{doerk2010atom, cote2000classical, cote2000ultracold, zhang2009near, zhang2009scattering, sayfutyarova2013charge, cote2002mesoscopic, schneider2010all, staanum2010rotational, roberts1998resonant, hudson2008inelastic, leo2000collision, leanhardt2003bose, lin2004impact}. Furthermore, these interactions are crucial for predicting ultra-cold two- and three-body scattering rates, lengths that directly relate to the stability and structure of Bose-Einstein condensates and the interaction of trapped species in optical and magnetic traps~\cite{PhysRevLett.88.040401}. The realization of cold-atom-ion hybrid traps has made the study of long-range interactions between different atomic systems even more vital. Therefore, accurate knowledge of these interactions and interactive potentials between various atomic and ionic species is required. Although experiments for investigating such potentials are quite elusive, theoretical interpretation of these interactions can assist in the simulation and interpretation of related experiments.

Due to their extraordinary electronic structures, elements of Groups I, II, and XII are found to be the most suitable for hybrid and optical traps \cite{tomza2019cold,smialkowski2020,wang2007,brickman2007,hachisu2008trapping}. 
Consequently, long-range interactions of these elements have been extensively studied~\cite{PhysRevA.101.032702,PhysRevA.81.062506,KAR201334,Shan-Shan.Lu.23202}, particularly in the context of magneto-optical traps (MOTs), trap losses, and ultra-cold collisions \cite{PhysRevA.65.041401, doi:10.1063/1.3691891, derevianko2001high, porsev2002high, PhysRevB.72.033109,shukla2022two}. Recently, transition-metal atoms, such as copper (Cu) and silver (Ag), have been successfully trapped~\cite{ PhysRevLett.101.103002,PhysRevA.62.063404}  {demonstrating} that hybrid trapping of these larger and denser atoms (Cu and Ag) with favorable elements of Groups I, II and XII is feasible. 
 {Because of their d-electrons, these metals exhibit stronger dispersion forces, which can lead to unusual interaction potentials when paired with other atoms or molecules. 
Understanding these interactions is crucial for designing hybrid quantum systems, where metallic atoms or nanoparticles might be used in conjunction with ultracold atomic gases. For example, Cu and Ag atoms might become part of surface traps or can be used in coupling schemes for quantum information or optical communication~\cite{henkel1999loss,tame2013quantum}.
In these applications, the interactive two- and three-body dispersion coefficients affect the strength and behavior of the interaction potential between two different atomic systems, which determines whether the atoms remain effectively bound at low temperatures, where thermal energy is not sufficient to overcome the interaction.}
Therefore, in the present work, we  {have studied} long-range interactions for various combinations of elements from Groups I, II, and XII with Cu and Ag, particularly reporting the two-body ($C_6$) and three-body ($C_9$) dispersion coefficients for the considered combinations.

The dispersion coefficients are calculated through the dynamic polarizabilities of the considered elements using a sum-over-states approach. We used experimental energies wherever possible to increase the accuracy of our calculation. The wave functions required for these polarizabilities are obtained using relativistic theoretical methods, since the considered elements possess strong relativistic effects. Using these polarizabilities, the dispersion coefficients are calculated for various atom-atom and atom-ion combinations. We also report the oscillator strengths and static dipole polarizabilities for the leading transitions of the considered elements of group XII.

The present manuscript is organized as follows: In Sec.~\ref{th} we provide a brief overview of the theory employed for the calculation of dispersion coefficients and dynamic dipole polarizability. The relativistic methods used to evaluate the wave functions of the atomic states are described in Sec.~\ref{methods} while Sec.~\ref{results}, discusses the evaluated results. Conclusions are drawn in Sec.~\ref{con}. All results are provided in atomic units (a.u.) unless otherwise stated.

\section{Theory}\label{th}
\subsection{vdW dispersion coefficients}
 {The attractive potential resulting from vdW interactions between two atoms, A and B in their ground states, is expressed as:}~\cite{reilly2015van}
\begin{equation}\label{eq1}
E_{AB} = -\frac{C_6}{R^{6}_{AB}}-\frac{C_{8}}{R^{8}_{AB}}-\frac{C_{10}}{R^{10}_{AB}}-...
\end{equation}
with $R_{AB}$ as the separation between the two atoms, $C_i$'s are the dispersion coefficients {arising} from instantaneous dipole-dipole ($C_6$), dipole-quadrupole ($C_8$), dipole-octupole ($C_{10}$) and {other interactions}. 
{Among all, the dipole-dipole interaction contributes the most to the long-range vdW interaction potential, and is given by:}
\begin{equation}
    C_6 = \frac{3}{\pi}\int_0^\infty \alpha^A(\iota\omega) \alpha^B(\iota\omega) d\omega
\end{equation}
where, $\alpha^A(\iota\omega)$ and $\alpha^B(\iota\omega)$ are the dynamic electric {dipole} polarizabilities of the considered atoms {A and B, respectively, at imaginary frequencies.}
However, when three atomic systems are involved, a three-body term is included to the pairwise additive expression of Eq.~\ref{eq1}. This higher order contribution beyond the pairwise approach of vdW potential is given as~\cite{reilly2015van} 
\begin{equation}
E_{ABC} = \frac{(3 \cos\theta_A \cos \theta_B \cos\theta_C +1)C_9}{(R_{AB}R_{BC}R_{CA})^3},
\end{equation} 
where, $\theta_A$, $\theta_B$, and $\theta_C$ are the angles formed between atoms such that $\theta_k = - \hat{R}_{ik}.\hat{R}_{kj}$. Depending on the angle between these atoms, this term can  {either be} positive or negative. {The corresponding} three-dipole vdW coefficient is given by 
\begin{equation}
  C_9 = \frac{3}{\pi}\int_0^\infty \alpha^A(\iota\omega)  \alpha^B(\iota\omega) \alpha^C(\iota\omega) d\omega,
\end{equation}
where, $\alpha^C(\iota\omega)$ is now imaginary frequency dependent polarizability of atom C.
 {Eventually, the accurate calculations of the dispersion coefficients depend on the precise evaluation of the polarizability, which is discussed in the next subsection.} 

\subsection{Dipole polarizability}
The general expression of dynamic dipole polarizability at imaginary frequency {for} any atomic system in its ground state {can be formulated as {\cite{10.1063/1.2841470}}:} 
\begin{equation}\label{alpha}
\alpha_v(\iota\omega) =  \sum_{k\neq v}\frac{f_{vk}}{(E_v-E_k)^2 + \omega^2}.
\end{equation} 
Here, $f_{vk}$ represents the oscillator strength of the dipole transitions from states $v$ to $k$ and $E_i$'s are the energies of the corresponding states.
This can further be expressed in terms of dipole matrix elements such that {\cite{10.1063/1.2841470}}
\begin{equation}\label{os}
f_{vk} = \frac{303.756}{g_v\lambda}|\langle\psi_k||\textbf{D}||\psi_v\rangle|^2.
\end{equation}
Here, $\langle\psi_k||\textbf{D}||\psi_v\rangle$ is the reduced {electric-}dipole (E1) matrix element with \textbf{D} as the  {electric-}dipole operator, $g_v$ {is the degeneracy (2$j_v$+1) of the lower state}, and $\lambda$ is the transition wavelength in angstroms.
Further, the dynamic dipole polarizability for an atomic system in its ground state at the imaginary frequency ($\alpha_v(\iota\omega)$), can be divided into the contributions as given below \cite{arora2012multipolar}: 
\begin{equation}
    \alpha_v(\iota\omega) = \alpha_{v,c}(\iota\omega) + \alpha_{v,val}(\iota\omega),
\end{equation}
where, $\alpha_{v,c}(\iota\omega)$ and  {$\alpha_{v,val} (\iota\omega)$} denote the contributions from core and valence correlations, respectively.  {These contributions can be evaluated independently using sum-over-states approach.}   {The core contribution arises due to the transition of core electrons to the bound intermediate states of an element.} Hence, the summation in Eq.~\ref{alpha} is taken for all the core orbitals to intermediate bound states.  
The valence contribution arises due to the transition of valence electrons to the higher unoccupied bound states.  {Further, this contribution can also be divided into two parts: main and tail.} The main part of the valence contribution is the most significant, and the summation in Eq.~\ref{alpha}  {is restricted}  over intermediate states for the low-lying transitions whereas the same summation is taken beyond the considered low{-}lying transitions to the bound high{-}lying transitions {for the calculation of} tail part. Further details regarding the polarizability calculations are available in \cite{shukla2020two,arora2007magic}. 

\section{Methods of Evaluation}\label{methods}
 {In order to evaluate the oscillator strengths of the dipole transitions, 
 precise computation of wave functions of the involved states is evident.}
 {We have implemented different methods for the computation of wave function required for evaluation of various contributions of dipole polarizability which are given as follows. 

\subsection{Dirac-Fock (DF) method}
 DF method has been used to compute the wave functions required for the evaluation of the core contributions and the tail part of valence contribution of dipole polarizability. This is an approximate method, employed because the core and tail contributions have minimal impact on the final polarizability results~\cite{PhysRevA.60.4476}.} We have implemented this method for the evaluation of DF state functions of all the elements considered in this work.
We begin with one electron DF equation given by~\cite{johnson2007atomic} 
\begin{equation}
    h(\textbf{r})\phi(\textbf{r})=\epsilon \phi(\textbf{r})
\end{equation}
 where 
\begin{equation}\label{DF2}
    h(\textbf{r}) = c \boldsymbol{\alpha} \cdot \textbf{p} + \boldsymbol{\beta} m c^2 - \frac{Z}{r} + U(\textbf{r})
\end{equation}
{In above equations,} the quantities $\boldsymbol{\alpha}$ and $\boldsymbol{\beta}$ are $4 \times 4$ Dirac matrices, $\textbf{p}$ is the momentum operator, $c$, $m$, $Z$ are the speed of light, mass of an electron and atomic number of the considered atom, respectively. $\phi$ is the one-electron orbital that describes the motion of an electron.
For many body problem, the complete DF Hamiltonian can be written in terms of one particle Hamiltonian $h(\textbf{r})$ as
\begin{equation}
    H_0 = \sum_i h(\textbf{r}_i). 
\end{equation}
In the second quantization, the Hamiltonian ($H_0$) is given in a.u. by 
\begin{equation}
 H_0=\sum_i \epsilon_ia_i^{\dagger} a_i \label{nopair},
\end{equation}
where the sum over $i$ include all electron orbitals and $\epsilon_i$ represents the {eigenvalue} of the one-electron DF orbitals. Consequently, the complete DF wave function for an atom is expressed in terms of Slater Determinant. 
The working DF wave functions {($|\Phi_v \rangle$)} {for} the interested states of the considered elements can be obtained by acting creation operator $a_v^{\dagger}$ on $|\Phi_0 \rangle$, such that, $|\Phi_v \rangle=a_v^{\dag} |\Phi_0 \rangle$.
 
\subsection{Relativistic Many-body perturbation theory approach}
We have used this method to compute the wave functions of Cu and Ag atoms.
The accurate computation of atomic wave functions encounters a major problem due to the presence of two-body electromagnetic interactions among electrons within the atomic system. In any typical approach, these atomic wave functions are obtained by using a mean-field approach and then incorporating the electron
correlation effects systematically.

In relativistic {many-body}  perturbation (RMBPT) approach, starting with the evaluation of {mean-field wave function ($|\Phi_0 \rangle$) }using the DF method, the atomic Hamiltonian is expressed as sum of Dirac-Fock Hamiltonian ($H_0$) and residual interaction ($V_I$) as,
\begin{equation}
H=H_0+V_I.
\end{equation}
Here, $H_0$ is the DF Hamiltonian given by Eq.~\ref{nopair} and $V_I$ is residual interaction expressed in a.u. as 
\begin{equation} \label{int}
V_I=\frac{1}{2}\sum_{ijkl} g_{ijkl}a_i^{\dag}a_j^{\dag}a_la_k - \sum_{ij} u_{ij}^{DF} a_i^{\dag} a_j,
\end{equation} 
where the sums over $i,j,k$ and $l$ include all electron orbitals and $g_{ijkl}$ is 
the two-body matrix element of the Coulomb interaction ($\frac{1}{r_{ij}}$) and $u_{ij}^{DF}$ denotes the DF potential~\cite{blundell1987formulas,johnson1996transition, safronova2005excitation}.  The choice of this $V^{N-1}$ DF potential (with $N$ number of electrons {in} the system) is made to facilitate the calculation of as many as states as possible that have a common closed core.

The corrections to the DF wave functions due to electron correlation effects are estimated by using the perturbative analysis of $V_I$. This is done by expressing the exact wave function of the state ($|\psi_v \rangle$) in the RMBPT analysis as
\begin{eqnarray}
 |\psi_v \rangle_{RMBPT} = |\phi_v \rangle + |\phi_v^{(1)} \rangle + |\phi_v^{(2)} \rangle + \cdots 
\end{eqnarray}
and its energy $(E_v$) as
\begin{eqnarray}
 E_v = E_v^{(0)} + E_v^{(1)}+ E_v^{(2)} + \cdots.
\end{eqnarray}
Here, superscripts $(0),(1),(2),$ etc., denote the order of perturbation and the zeroth-order energy is $E_v^{(0)}= \sum_p^N \epsilon_p$. After obtaining the wave functions, the electric-dipole matrix element $(D_{vk})$ between the states $|\Psi_v \rangle$ and $|\Psi_k \rangle$ is calculated as follows: 
\begin{equation}
D_{vk}=\frac{\left\langle\Psi_v|\textbf{D}|\Psi_k\right\rangle}{\sqrt{\left\langle\Psi_k|\Psi_k\right\rangle \left\langle\Psi_v|\Psi_v\right\rangle}},  \label{matel}
\end{equation}
where $ \textbf{D}$ is the electric-dipole operator of the considered transition. Additionally, these matrix elements can be reduced in the form $\langle J_v||D||J_k\rangle$, using Wigner--Eckart theorem as~\cite{lindgren2012atomic}
\begin{equation}
D_{vk}=(-1)^{J_k-M_{J_k}}\left(\begin{array}{ccc}
J_k & 1 & J_v\\
-M_{J_k} & 0 & M_{J_v}
\end{array}\right) \langle J_v||D||J_k\rangle,
\end{equation}
 {where, the term in braces denotes the 3-j symbol.}

\subsection{Relativistic all-order approach}
For monovalent systems, that is, atoms of group I and singly charged ions of group II and XII, we computed the atomic {wave functions} by using the relativistic all-order method. After evaluating the DF wave function, the exact wave function {is constructed} by using single-double approximation in the all-order method, as given in the following expression~\cite{PhysRevA.60.4476}:
\begin{equation}
\begin{aligned}
 |\psi _v \rangle _{SD} = \Bigg[1 + \sum_{ma} \rho_{ma} a^{\dagger}_m a_a + \frac{1}{2} \sum_{mnab} \rho_{mnab} a^\dag_m a^\dag_n a_b a_a \\ + \sum_{m\ne v} \rho_{mv} a^\dag_m a_v + \sum_{mna} \rho_{mnva} a^\dag_m a^\dag_n  a_a a_v \bigg] |\Phi _v \rangle,
\end{aligned}
\end{equation}
where indices $\{a,b\}$ and $\{m,n\}$ denote the core and virtual orbitals of the DF wave functions $|\Phi _v \rangle$ of the state $v$. The terms $\rho_{ma}$ and $\rho_{mnab}$ represent the single and double  {excitation} coefficients  {whereas} $a_i$ and $a^\dag_i$ denote the annihilation and creation operators. This wave function  {is further augmented} by including the missing important triple excitation coefficients,  {thus defining the final wave function ($|\psi_v\rangle_{SDpT}$) as~\cite{PhysRevA.60.4476}}
\begin{eqnarray}		
|\psi_v\rangle_{SDpT} &=& |\psi_v\rangle_{SD} + \left[ \frac{1}{18} \sum_{mnlabc} \rho_{mnlabc} a_m^\dagger a_n^\dagger a_l^\dagger a_c a_b a_a \right.
						\nonumber\\
& & \left. + \frac{1}{6} \sum_{mnlab} \rho_{mnlvab} a_m^\dagger a_n^\dagger a_l^\dagger a_b a_a a_v \right]|\Phi_v \rangle.   
	 \end{eqnarray} 

\begin{table*}
\caption{\label{table1} Contributions to the ground state dipole polarizabilities (in a.u.) of the Cu and Ag atoms  {are presented here}. Various contributions{, including} the oscillator strength ($f_{mn}$) contributing to the main part of the valence correlations, are explicitly {quoted}. Tail and core contributions are also provided in this table. {The} final results are compared with the previously available values. 
}
	\begin{center}
\begin{tabular}{cccccccccc}
\hline
\hline
\multicolumn{4}{c}{Cu}  & \multicolumn{4}{c}{Ag}\\[1.5ex]
\hline\\
Contribution  & \multicolumn{2}{c}{$f_{vk}$}  & $\alpha(0)$  & Contribution  & \multicolumn{2}{c}{$f_{vk}$}  & $\alpha(0)$	&   \\ [1.5ex]
& Present & NIST &&& Present & NIST &\\
\hline\\[0.15ex]
$4S_{1/2}$	-	$4P_{1/2}$	&	2.585[-1] & 2.212[-1] &	13.36  & 	$5S_{1/2}$	-	$5P_{1/2}$ 	&	2.343[-1] & 2.2[-1] &	12.92 &\\[0.25ex]
$4S_{1/2}$	-	$5P_{1/2}$	&	4.428[-3] & 6.0[-3]	&	0.09		& $5S_{1/2}$	-	$6P_{1/2}$	&	2.403[-3] &9.6[-4]	&	0.05  &\\[0.25ex]
$4S_{1/2}$	-	$6P_{1/2}$	&	6.620[-4] &	&	0.01		& $5S_{1/2}$	-	$7P_{1/2}$	&	2.393[-4] &	&	0.004 &	\\[0.25ex]
$4S_{1/2}$	-	$7P_{1/2}$	&	2.735[-4]&	&	0.004	&	$5S_{1/2}$	-	$8P_{1/2}$	&	5.375[-5]&	&	0.0008	 & \\[0.25ex]
$4S_{1/2}$	-	$8P_{1/2}$	&	3.252[-5]&	&	$\sim0$	  	&	$5S_{1/2}$	-	$9P_{1/2}$	&	3.171[-6]&	&	$\sim0$	&  \\	[0.25ex]
$4S_{1/2}$	-	$4P_{3/2}$	&	5.20[-1]& 4.41[-1]	&	26.43		&	$5S_{1/2}$	-	$5P_{3/2}$	&	4.802[-1]& 4.5[-1]	&	24.91 & \\[0.25ex]
$4S_{1/2}$	-	$5P_{3/2}$	&	1.016[-2]&1.2[-2]	&	0.20	 	&	$5S_{1/2}$	-	$6P_{3/2}$	&	8.717[-3]& 4.0[-3]	&	0.178 &\\[0.25ex]
$4S_{1/2}$	-	$6P_{3/2}$	&	1.654[-3]&	&	0.026		&	$5S_{1/2}$	-	$7P_{3/2}$	&	1.411[-3]	&	& 0.023 &  \\[0.25ex]
$4S_{1/2}$	-	$7P_{3/2}$	&	7.612[-4]&	&	0.01		& $5S_{1/2}$	-	$8P_{3/2}$	&	6.664[-4]	&	& 0.01  & \\[0.25ex]
$4S_{1/2}$	-	$8P_{3/2}$	&	1.44[-4]&	&	0.002		& $5S_{1/2}$	-	$9P_{3/2}$	&	1.629[-4]	&	& 0.002 &		\\[0.25ex]
Main	&		&&	40.13	&	 Main	&	&	&	38.10	 &	\\[0.5ex]	
  Tail & & & 0.15  	& Tail & & & 0.22 &\\[0.5ex]
 Core & & & 4.404  	& Core & && 8.873  & \\[0.5ex]
  Total && &  44.68  	 & Total & && 47.20 & \\[0.5ex]
 \hline\hline
    \end{tabular}	   
	 \end{center}
	 \end{table*}

\begin{table}
\caption{\label{pd} Percentage Deviation of our calculated values of dipole polarizability with available theoretical and experimental values.}
\label{pd}
\begin{center}
\begin{tabular}{p{4.8cm}p{0.8cm}p{0.8cm}p{1.0cm}p{0.8cm}}
 \hline
 \multicolumn{1}{c}{Method} & \multicolumn{2}{c}{Cu} & \multicolumn{2}{c}{Ag}  \\ 
 \hline\\
   & $\alpha(0)$ & $\delta(\%)$ & $\alpha(0)$ & $\delta(\%)$ \\  [0.5ex]
Theoretical Values & & & & \\
\hline
Present & 44.68 & & 47.20 & \\[0.5ex]
CCSD(T)+ NENpPolMe~\cite{ijqc} & 46.50 & 3.9  & 52.50 & 10 \\[0.5ex]
 SE Hamiltonian~\cite{PhysRevA.78.062710} & 41.65 & 2.3 & 46.17 & 2.2 \\[0.5ex]
  TDDFT~\cite{gould2016c} & 41.70 & 7.1 & 46.20 & 2.2 \\ [0.5ex]
 CCSD(T)+ NENpPolMe+EE~\cite{smialkowski2021highly} & 45.9 & 2.7 & 50.10 & 5.8\\ [0.5ex]
 \\
Experimental Values & & & &\\
\hline 

 {LI+IP technique}~\cite{10.1063/5.0084981} & $35\pm7$  & & $56\pm14$ &  \\
 \hline\hline
\end{tabular}
\end{center}
 \end{table}
  
\subsection{Multi-configuration Dirac-Fock (MCDF) approach}
The desired wave functions for the divalent systems, i.e., for group II and XII atoms, are evaluated by employing MCDF approach using the GRASP2K code \cite{jonsson2013new}. In this approach, an atomic state wave function (ASF) is expressed as a linear combination of possible configurational state functions (CSFs) with the same parity and total angular momentum as that of ASF,  {given by}
 \begin{equation}
     |\psi_v \rangle_{MCDF} = \sum_{c = 1}^N a_c |\phi_c \rangle .
 \end{equation}
Here, the variable $c$ denotes the number of CSFs ($|\phi_c \rangle$), and $a_c$ represents the mixing coefficient for the  CSFs. Initially, the single-particle orbital radial functions are calculated numerically using the Dirac-Coulomb Hamiltonian through a multi-configuration self-consistent-field method. Relativistic configuration-interaction calculations are then performed to incorporate both, Breit as well as quantum electrodynamics (QED) corrections, thereby, considering a large number of CSFs, especially with mixing coefficients $> 10^{-3}$, for the computation of precise ASFs.

 {Consequently, the matrix elements are obtained using Eq.~\ref{matel} by employing the computed ASFs. These matrix elements are further used to evaluate oscillator strengths using Eq.~\ref{os} for each of the considered transitions required for the evaluation of main part of the dipole polarizability.} 

\section{Results and Discussion}\label{results}
In the present work, we report static and dynamic polarizabilities of $4S_{1/2}$ and $5S_{1/2}$ states of Cu and Ag, respectively, {and their vdW interaction coefficients with the elements of group I, II and XII. We have also presented an explicit comparison of our obtained results with the available literature for ascertaining the accuracy of our work. We have not given the results of static dipole polarizability of atoms of group I, II and ions of group II and XII as it is already reported by us in previous works \cite{UDportal,kaur2021dispersion,shukla2020two}. Further discussion of the calculated results is provided in their relevant subsections.}

\subsection{Dipole  Polarizability}
Table~\ref{table1} demonstrates the results for static dipole polarizabilities ($\alpha(0)$) of $4S_{1/2}$ and $5S_{1/2}$ states of Cu and Ag atoms, respectively, along with the contributions from various dominant transitions and their corresponding oscillator strengths. As expected, the oscillator strengths of the contributing transitions decrease with an increase in principal quantum number $n$. We have also observed a good accord among the oscillator strengths of $4S_{1/2} \rightarrow 4P_{1/2,3/2}$ for Cu and $5S_{1/2} \rightarrow 5P_{1/2,3/2}$ for Ag with respect to the oscillator strength provided in NIST database~\cite{NIST_ASD}.
However, the deviation of the values of oscillator strength from the reported values is observed to increase for less dominant transitions where the $\alpha(0)$ contribution is small. It is to be noted that the evaluation of oscillator strengths provided on NIST database is done using theoretical calculations of lifetimes of different states without incorporating higher-order relativistic corrections leading to a larger deviation in oscillator strengths with involved $4P$ states of Cu. 
The evaluated oscillator strengths are further used in the determination of main component of the static dipole plarizabilities of ground states of Cu and Ag atoms. The main, tail and core contributions are eventually added to get the final value of static polarizability as given in Table~\ref{table1}. Furthermore, in Table~\ref{pd}, we have compared our results for $\alpha(0)$ values for the considered atoms with available literature reported by Neogrady \textit{et al.}\cite{ijqc}, Zhang \textit{et al.}\cite{PhysRevA.78.062710}, Gould \textit{et al.}\cite{gould2016c} and Smialkowski \textit{et al.}\cite{smialkowski2021highly} who have calculated the 
$\alpha(0)$ values by employing coupled-cluster Douglas-Kroll no-pair formalism (CCSD(T)+ NENpPolMe)~\cite{ijqc}, a semiempirical method in a large-dimension single-electron basis~\cite{PhysRevA.78.062710}, Time Dependent Density Funtional Theory (TDDFT)~\cite{gould2016c} and CCSD(T)+ NENpPolMe method using experimental energies (EE)~\cite{smialkowski2021highly}, respectively.
The percentage deviation of our calculated values from the reported values is also mentioned in the same table. As can be seen from Table~\ref{pd}, in general, less than $10\%$ of deviation is seen among theoretical values of $\alpha(0)$ reported in various studies.
These discrepancies can usually be attributed to the different methods employed in the respective theoretical approaches. In addition, the low discrepancies of $\sim 2\%$ in $\alpha(0)$ w.r.t. Zhang \textit{et al.}~\cite{PhysRevA.78.062710} is expected due to the fact that Zhang \textit{et al.} have not considered the spin-orbit splitting among the atomic levels, which further leads to the underlying uncertainties of $8\%$ and $5\%$ in oscillator strengths of Cu and Ag, respectively and an overall $10\%$ uncertainty in their core polarizability values in their own work. In contrast, the deviation of static dipole polarizabilities from the results provided by Neogrady \textit{et al.} and Smialkowski \textit{et al.}\cite{smialkowski2021highly} can be advocated by the fact that these studies are based on the \textit{ab initio} approaches with large basis sets and small-core relativistic energy-consistent pseudo-potentials, that only account for scalar relativistic effects. The negligence of spin-orbit coupling and other vector relativistic effects leads to the inherent deviation among the reported results.
Our values show a  {reasonable variation of $7.1\%$ and $2.2\%$ corresponding to Cu and Ag, in comparison to} the available TDDFT results\cite{gould2016c}.  We consider our values to be more reliable compared to the TDDFT results due to the use of a sum-over-states approach and accurate calculations of oscillator strengths. Moreover, strongly correlated systems are poorly treated using the TDDFT approach and hence, is not believed to be a good scheme for considering excitations~\cite{burke2012perspective}.
Overall, our calculated values are relevant and exhibit good agreement with previously reported theoretical calculations~\cite{ijqc,gould2016c,PhysRevA.78.062710,smialkowski2021highly} and experimental~\cite{PhysRevA.91.010501} values with less than 10$\%$ variation. Hence, our further evaluations of the $C_6$ and $C_9$ dispersion coefficients, which are derived using these dipole polarizability, can be considered reliable.

\begin{figure}
    \includegraphics[width=\columnwidth,keepaspectratio]{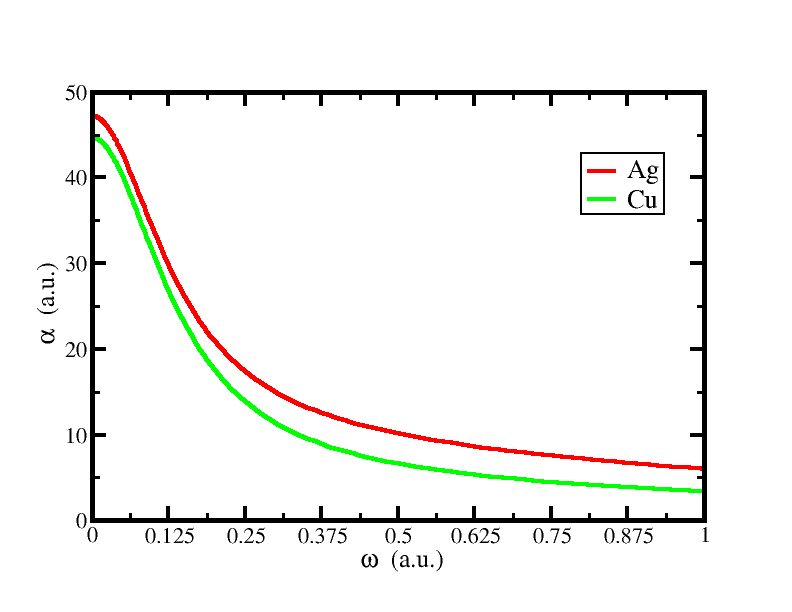}
\caption{Dynamic dipole polarizability $\alpha$ (in a.u.) at imaginary frequency of Cu and Ag.}
\label{group11}
\end{figure}

By using the calculated oscillator strengths, we evaluated the dynamic dipole polarizability of Cu and Ag atoms 
at imaginary frequencies. The plots of dynamic dipole polarizability ($\alpha (\iota\omega)$) {versus} the imaginary frequency ($\omega$) for Cu and Ag  are presented in Fig~\ref{group11}.  From the figure, we can observe that as the frequency increases, the dipole polarizability of the atoms starts to decrease. The polarizability for Cu beyond $\omega > 0.17$ a.u. decreases more rapidly as compared to Ag. The plots of dynamic polarizability of group I atoms, group II and XII atoms and ions are given in our previous works~\cite{UDportal,kaur2021dispersion,shukla2020two}.  

\subsection{Dispersion Coefficients}
\begin{table}
\caption{{$C_6$} Dispersion Coefficients for Cu and Ag atoms {interacting} with {atoms of} group I (Li, Na, K, Rb, Cs, Fr), group II (Be, Mg, Ca, Sr, Ba) and group XII  (Zn, Cd, Hg) and their comparison with previously reported results.}
\label{c6}
\begin{center}
\begin{tabular}{p{0.5cm}p{1cm}p{1cm}p{1cm}p{1cm}p{1cm}p{1cm}}
 \hline
 \hline
 \multicolumn{1}{c}{Element} & \multicolumn{3}{c}{Cu} & \multicolumn{3}{c}{Ag} \\
 \hline\\[0.015ex]
   & Present & Other & $\delta(\%)$ & Present & Other &$\delta(\%)$ \\ 
 \hline\\[0.015ex]
 Li & 515 & 523$^a$ & 1.52 & 562 & 567$^a$ & 0.88\\ [0.5ex]

 Na & 560 & 564$^a$ & 0.71 & 616 & 611$^a$ & 0.82\\ [0.5ex]
 K & 862 & 864$^a$ & 0.23 & 951 & 937$^a$ & 1.49\\[0.5ex]
 Rb & 957 & 950$^a$ & 0.73 & 1061 & 1031$^a$ & 2.91\\ [0.5ex]
 Cs & 1161 & 1138$^a$ & 2.02 & 1294 & 1234$^a$ & 4.86 \\ [0.5ex]
 Fr & 1076 & 1030$^a$ & 4.47 & 1214 & 1116$^a$ & 8.78\\ [0.5ex]
 \hline\\[0.015ex]
  Be$^+$ & 128 && & 145 & & \\ [0.5ex]
  Mg$^+$ & 193 &&  & 220 & & \\ [0.5ex]
  Ca$^+$ & 363 & &  & 412 & &  \\ [0.5ex]
  Sr$^+$ & 441 & && 504 & &\\ [0.5ex]
  Ba$^+$ & 581 & && 667 &  &\\ [0.5ex]
\hline \\ [0.015ex]
  Be & 221 & 214$^a$ & 3.27 & 252 & 231$^a$ & 9.09 \\ [0.5ex]
  Mg & 399 & 371$^a$ & 7.55 & 451 & 400$^a$ & 12.75  \\ [0.5ex]
  Ca & 720 & 681$^a$ & 5.73& 807 & 737$^a$  & 9.50\\ [0.5ex]
  Sr & 938 & 821$^a$ & 14.25 & 1049 & 889$^a$ & 18.00  \\ [0.5ex]
  Ba & 1123 & 1049$^a$ & 7.05 & 1257 & 1136$^a$ & 10.65  \\ [0.5ex]
\hline \\ [0.015ex]
 Cu & 253 & 264$^b$ & 4.17 & 293 & & 	\\ [0.5ex]
 Ag & 293 & &  	& 348 & 341$^b$ & 2.05 \\ [0.5ex]
 \hline\\[0.015ex]
  Zn$^+$ & 129 && & 154 &  & \\ [0.5ex]
  Cd$^+$ & 179 &&  & 217 & & \\ [0.5ex]
  Hg$^+$ & 176 & & & 221 & & \\  [0.5ex]
 \hline\\[0.015ex]
  Zn & 238 &&  & 279 & & \\ [0.5ex]
  Cd & 351 &&  & 416 & &\\ [0.5ex]
  Hg & 241 &&-& 296 &  &\\  [0.5ex]
 \hline\hline
\end{tabular}
\end{center}
$^a$ Ref \cite{smialkowski2021highly} (CCSD(T)+ NENpPolMe+EE),  $^b$Ref\cite{gould2016c} (TDDFT)\\
 \end{table}
Using the calculated dynamic dipole polarizability values, we obtain the $C_6$ dispersion coefficients for the combinations of two atoms, i.e., Cu and Ag with elements from the group I, II and XII. $C_6$ coefficient values for heteronuclear and homonuclear dimers are listed in Table~\ref{c6}. 
In this table, we also present a comparison of our results with the previously reported values, wherever possible.
The purpose of this comparison is to ascertain the reliability of the method used for our calculations. 
On comparison, we find that our results are in good agreement with the results given by Simalkowski \textit{et al.}~\cite{smialkowski2021highly} and Gould \textit{et al.}~\cite{gould2016c}. Our {calculated}  $C_6$ coefficients for Cu and Ag atoms with group I and II atoms differs by $\sim2.5\%\ $ and $\sim10.5\%\ $ from the reported results, respectively. The $C_6$ coefficients for various combinations of Cu and Ag atoms with themselves (except for Cu and Ag dimers) and group XII atoms have not been previously reported in the literature.
From Table~\ref{c6}, we observe specific trends in the relative magnitudes of the calculated $C_6$ coefficients for interactions of Cu and Ag atoms with the atoms of group I, II, and XII. The Ag-Cs interaction exhibits the highest $C_6$ coefficient, while the Cu-Be$^+$ and Cu-Zn$^+$ interactions demonstrate the lowest coefficient values among all the other combinations.
  \begin{table}
  \caption{$C_9$ Dispersion Coefficients for Cu and Ag atoms interacting with atoms of group I (Li, Na, K, Rb, Cs, Fr), group II (Be, Mg, Ca, Sr, Ba) and Group XII (Zn, Cd, Hg).}
\label{c9}
\begin{center}
 \begin{tabular}{c@{\hskip 0.5in}c@{\hskip 0.5in}c}
 \hline
 \hline\\[0.2ex]
 Atom  &  Cu-Cu  &  Ag-Ag   \\ [0.5ex] 
 \hline\\[0.015ex]
 Li & 19008 & 21947  \\ [0.5ex] 
 Na & 20237 & 23464 \\ [0.5ex] 
 K & 31504 & 36430 \\ [0.5ex] 
 Rb& 34546 & 40044 \\ [0.5ex] 
 Cs  & 41601 & 48296  \\ [0.5ex] 
 Fr & 36977 & 43282 \\ [0.5ex] 
 \hline \\ [0.015ex]
  Be$^+$  & 4121  & 4893   \\ [0.5ex] 
  Mg$^+$  & 6093  & 7261  \\  [0.5ex] 
  Ca$^+$  & 11784 & 13939 \\ [0.5ex] 
  Sr$^+$  & 14155 & 15414 \\ [0.5ex] 
  Ba$^+$  & 18582 & 22050 \\ [0.5ex] 
\hline \\[0.015ex]
  Be & 6890 & 8258 \\ [0.5ex] 
  Mg  & 12793 & 15221\\ [0.5ex] 
  Ca  & 24375 & 28644 \\ [0.5ex] 
  Sr  & 31956 & 37491 \\ [0.5ex] 
  Ba  & 38591 & 45125 \\ [0.5ex] 
\hline \\[0.015ex]
  Cu  & 7666 & 9215 \\ [0.5ex] 
  Ag  & 8382 & 10205 \\ [0.5ex] 
 \hline\\[0.015ex]
  Zn$^+$  & 3559 & 4365 \\ [0.5ex] 
  Cd$^+$  & 4685 & 5807 \\ [0.5ex] 
  Hg$^+$  & 4157 & 5286 \\ [0.5ex] 
 \hline\\[0.015ex]
  Zn  & 6962 & 8454 \\ [0.5ex] 
  Cd & 9925 & 12135 \\ [0.5ex] 
  Hg & 6148 & 7689 \\ [0.5ex] 
 \hline
 \hline
 
\end{tabular}
\end{center}
\end{table}

In addition, we have calculated the $C_9$ coefficients for the atom-atom-atom combinations, where two atoms are from group XI, and the third atom/ion is from group I, II, XII or from among the Cu and Ag atoms. These coefficients are listed in Table~\ref{c9}. From this table, we observe that the relative magnitudes of the calculated $C_9$ coefficients follow the same trend as that of $C_6$ values, i.e., (Ag-Ag-X) $>$ (Cu-Cu-X) (where X refers to the atom of group I, II, XI, or XII). The strongest interaction is for Ag-Ag-Cs and lowest for Cu-Cu-Zn$^+$ and Cu-Cu-Be$^+$ as observed from the values of $C_9$ coefficients given in the table.
For all the considered combinations, there are no previously reported values.
 
\section{Conclusion}\label{con}
In the present work, we performed a detailed calculation of the $C_6$ and $C_9$ interaction coefficients for Cu and Ag atoms with atoms from group I  (Li, Na, K, Rb, Cs, Fr), group II (Be, Mg, Ca, Sr, Ba) and group XII (Zn, Cd, Hg) and singly charged ions of group II (Be$^+$, Mg$^+$, Ca$^+$, Sr$^+$, Ba$^+$) and group XII (Zn$^+$, Cd$^+$, Hg$^+$) using relativistic methods. For this purpose, the oscillator strength, static and dynamic dipole polarizabilty of these atoms are  also evaluated. Wherever possible, we have compared our results with the corresponding values from other reported measurements and theoretical studies. The overall good agreement provides confidence in the reliability of our reported values. We have also observed that the Ag-Cs interaction has the highest $C_6$ coefficients, while the  Cu-Be$^+$ interaction has the lowest among all the combinations. Similarly, for the $C_9$ coefficients, the Ag-Ag-Cs interaction demonstrates the highest value. 
 {We believe that our calculated results will encourage more experimental and theoretical studies in this direction and will be useful to researchers in various fields, specifically in the study of van der Waals complexes, ultracold physics, hybrid trapping and coupling schemes for different quantum processes.} 

\section*{Acknowledgement}
Research at Perimeter Institute is supported in part by the Government of Canada through the Department of Innovation, Science and Economic Development and by the Province of Ontario through the Ministry of Colleges and Universities.

\section*{Data Availability}
This manuscript has associated data in a data repository. [Authors’ comment: The underlying data in this manuscript is not currently available to the public, but can be available from the authors upon reasonable requests.]

\section*{Author's Contribution}
\textbf{Harpreet Kaur}: Conceptualized the study, performed the calculations, drafted the manuscript.\\
 \textbf{Jyoti}: Participated in the discussions of the results and manuscript.\\
 \textbf{Neelam Shukla} : Participated in the discussions of the results and manuscript. \\
 \textbf{Bindiya Arora}: Supervised the research, reviewed and edited the manuscript.

\bibliography{c6c9.bib}

\begin{thebibliography}{56}
\expandafter\ifx\csname natexlab\endcsname\relax\def\natexlab#1{#1}\fi
\expandafter\ifx\csname bibnamefont\endcsname\relax
  \def\bibnamefont#1{#1}\fi
\expandafter\ifx\csname bibfnamefont\endcsname\relax
  \def\bibfnamefont#1{#1}\fi
\expandafter\ifx\csname citenamefont\endcsname\relax
  \def\citenamefont#1{#1}\fi
\expandafter\ifx\csname url\endcsname\relax
  \def\url#1{\texttt{#1}}\fi
\expandafter\ifx\csname urlprefix\endcsname\relax\def\urlprefix{URL }\fi
\providecommand{\bibinfo}[2]{#2}
\providecommand{\eprint}[2][]{\url{#2}}

\bibitem[{\citenamefont{Doerk et~al.}(2010)\citenamefont{Doerk, Idziaszek, and Calarco}}]{doerk2010atom}
\bibinfo{author}{\bibfnamefont{H.}~\bibnamefont{Doerk}}, \bibinfo{author}{\bibfnamefont{Z.}~\bibnamefont{Idziaszek}}, \bibnamefont{and} \bibinfo{author}{\bibfnamefont{T.}~\bibnamefont{Calarco}}, \bibinfo{journal}{Physical Review A} \textbf{\bibinfo{volume}{81}}, \bibinfo{pages}{012708} (\bibinfo{year}{2010}).

\bibitem[{\citenamefont{C{\^o}t{\'e}}(2000)}]{cote2000classical}
\bibinfo{author}{\bibfnamefont{R.}~\bibnamefont{C{\^o}t{\'e}}}, \bibinfo{journal}{Physical review letters} \textbf{\bibinfo{volume}{85}}, \bibinfo{pages}{5316} (\bibinfo{year}{2000}).

\bibitem[{\citenamefont{C{\^o}t{\'e} and Dalgarno}(2000)}]{cote2000ultracold}
\bibinfo{author}{\bibfnamefont{R.}~\bibnamefont{C{\^o}t{\'e}}} \bibnamefont{and} \bibinfo{author}{\bibfnamefont{A.}~\bibnamefont{Dalgarno}}, \bibinfo{journal}{Physical Review A} \textbf{\bibinfo{volume}{62}}, \bibinfo{pages}{012709} (\bibinfo{year}{2000}).

\bibitem[{\citenamefont{Zhang et~al.}(2009{\natexlab{a}})\citenamefont{Zhang, Bodo, and Dalgarno}}]{zhang2009near}
\bibinfo{author}{\bibfnamefont{P.}~\bibnamefont{Zhang}}, \bibinfo{author}{\bibfnamefont{E.}~\bibnamefont{Bodo}}, \bibnamefont{and} \bibinfo{author}{\bibfnamefont{A.}~\bibnamefont{Dalgarno}}, \bibinfo{journal}{The Journal of Physical Chemistry A} \textbf{\bibinfo{volume}{113}}, \bibinfo{pages}{15085} (\bibinfo{year}{2009}{\natexlab{a}}).

\bibitem[{\citenamefont{Zhang et~al.}(2009{\natexlab{b}})\citenamefont{Zhang, Dalgarno, and C{\^o}t{\'e}}}]{zhang2009scattering}
\bibinfo{author}{\bibfnamefont{P.}~\bibnamefont{Zhang}}, \bibinfo{author}{\bibfnamefont{A.}~\bibnamefont{Dalgarno}}, \bibnamefont{and} \bibinfo{author}{\bibfnamefont{R.}~\bibnamefont{C{\^o}t{\'e}}}, \bibinfo{journal}{Physical Review A} \textbf{\bibinfo{volume}{80}}, \bibinfo{pages}{030703} (\bibinfo{year}{2009}{\natexlab{b}}).

\bibitem[{\citenamefont{Sayfutyarova et~al.}(2013)\citenamefont{Sayfutyarova, Buchachenko, Yakovleva, and Belyaev}}]{sayfutyarova2013charge}
\bibinfo{author}{\bibfnamefont{E.~R.} \bibnamefont{Sayfutyarova}}, \bibinfo{author}{\bibfnamefont{A.~A.} \bibnamefont{Buchachenko}}, \bibinfo{author}{\bibfnamefont{S.~A.} \bibnamefont{Yakovleva}}, \bibnamefont{and} \bibinfo{author}{\bibfnamefont{A.~K.} \bibnamefont{Belyaev}}, \bibinfo{journal}{Physical Review A} \textbf{\bibinfo{volume}{87}}, \bibinfo{pages}{052717} (\bibinfo{year}{2013}).

\bibitem[{\citenamefont{Cote et~al.}(2002)\citenamefont{Cote, Kharchenko, and Lukin}}]{cote2002mesoscopic}
\bibinfo{author}{\bibfnamefont{R.}~\bibnamefont{Cote}}, \bibinfo{author}{\bibfnamefont{V.}~\bibnamefont{Kharchenko}}, \bibnamefont{and} \bibinfo{author}{\bibfnamefont{M.}~\bibnamefont{Lukin}}, \bibinfo{journal}{Physical review letters} \textbf{\bibinfo{volume}{89}}, \bibinfo{pages}{093001} (\bibinfo{year}{2002}).

\bibitem[{\citenamefont{Schneider et~al.}(2010)\citenamefont{Schneider, Roth, Duncker, Ernsting, and Schiller}}]{schneider2010all}
\bibinfo{author}{\bibfnamefont{T.}~\bibnamefont{Schneider}}, \bibinfo{author}{\bibfnamefont{B.}~\bibnamefont{Roth}}, \bibinfo{author}{\bibfnamefont{H.}~\bibnamefont{Duncker}}, \bibinfo{author}{\bibfnamefont{I.}~\bibnamefont{Ernsting}}, \bibnamefont{and} \bibinfo{author}{\bibfnamefont{S.}~\bibnamefont{Schiller}}, \bibinfo{journal}{Nature Physics} \textbf{\bibinfo{volume}{6}}, \bibinfo{pages}{275} (\bibinfo{year}{2010}).

\bibitem[{\citenamefont{Staanum et~al.}(2010)\citenamefont{Staanum, H{\o}jbjerre, Skyt, Hansen, and Drewsen}}]{staanum2010rotational}
\bibinfo{author}{\bibfnamefont{P.~F.} \bibnamefont{Staanum}}, \bibinfo{author}{\bibfnamefont{K.}~\bibnamefont{H{\o}jbjerre}}, \bibinfo{author}{\bibfnamefont{P.~S.} \bibnamefont{Skyt}}, \bibinfo{author}{\bibfnamefont{A.~K.} \bibnamefont{Hansen}}, \bibnamefont{and} \bibinfo{author}{\bibfnamefont{M.}~\bibnamefont{Drewsen}}, \bibinfo{journal}{Nature Physics} \textbf{\bibinfo{volume}{6}}, \bibinfo{pages}{271} (\bibinfo{year}{2010}).

\bibitem[{\citenamefont{Roberts et~al.}(1998)\citenamefont{Roberts, Claussen, Burke~Jr, Greene, Cornell, and Wieman}}]{roberts1998resonant}
\bibinfo{author}{\bibfnamefont{J.}~\bibnamefont{Roberts}}, \bibinfo{author}{\bibfnamefont{N.}~\bibnamefont{Claussen}}, \bibinfo{author}{\bibfnamefont{J.~P.} \bibnamefont{Burke~Jr}}, \bibinfo{author}{\bibfnamefont{C.~H.} \bibnamefont{Greene}}, \bibinfo{author}{\bibfnamefont{E.~A.} \bibnamefont{Cornell}}, \bibnamefont{and} \bibinfo{author}{\bibfnamefont{C.}~\bibnamefont{Wieman}}, \bibinfo{journal}{Physical Review Letters} \textbf{\bibinfo{volume}{81}}, \bibinfo{pages}{5109} (\bibinfo{year}{1998}).

\bibitem[{\citenamefont{Hudson et~al.}(2008)\citenamefont{Hudson, Gilfoy, Kotochigova, Sage, and DeMille}}]{hudson2008inelastic}
\bibinfo{author}{\bibfnamefont{E.~R.} \bibnamefont{Hudson}}, \bibinfo{author}{\bibfnamefont{N.~B.} \bibnamefont{Gilfoy}}, \bibinfo{author}{\bibfnamefont{S.}~\bibnamefont{Kotochigova}}, \bibinfo{author}{\bibfnamefont{J.~M.} \bibnamefont{Sage}}, \bibnamefont{and} \bibinfo{author}{\bibfnamefont{D.}~\bibnamefont{DeMille}}, \bibinfo{journal}{Physical review letters} \textbf{\bibinfo{volume}{100}}, \bibinfo{pages}{203201} (\bibinfo{year}{2008}).

\bibitem[{\citenamefont{Leo et~al.}(2000)\citenamefont{Leo, Williams, and Julienne}}]{leo2000collision}
\bibinfo{author}{\bibfnamefont{P.~J.} \bibnamefont{Leo}}, \bibinfo{author}{\bibfnamefont{C.~J.} \bibnamefont{Williams}}, \bibnamefont{and} \bibinfo{author}{\bibfnamefont{P.~S.} \bibnamefont{Julienne}}, \bibinfo{journal}{Physical review letters} \textbf{\bibinfo{volume}{85}}, \bibinfo{pages}{2721} (\bibinfo{year}{2000}).

\bibitem[{\citenamefont{Leanhardt et~al.}(2003)\citenamefont{Leanhardt, Shin, Chikkatur, Kielpinski, Ketterle, and Pritchard}}]{leanhardt2003bose}
\bibinfo{author}{\bibfnamefont{A.}~\bibnamefont{Leanhardt}}, \bibinfo{author}{\bibfnamefont{Y.}~\bibnamefont{Shin}}, \bibinfo{author}{\bibfnamefont{A.}~\bibnamefont{Chikkatur}}, \bibinfo{author}{\bibfnamefont{D.}~\bibnamefont{Kielpinski}}, \bibinfo{author}{\bibfnamefont{W.}~\bibnamefont{Ketterle}}, \bibnamefont{and} \bibinfo{author}{\bibfnamefont{D.}~\bibnamefont{Pritchard}}, \bibinfo{journal}{Physical review letters} \textbf{\bibinfo{volume}{90}}, \bibinfo{pages}{100404} (\bibinfo{year}{2003}).

\bibitem[{\citenamefont{Lin et~al.}(2004)\citenamefont{Lin, Teper, Chin, and Vuleti{\'c}}}]{lin2004impact}
\bibinfo{author}{\bibfnamefont{Y.-j.} \bibnamefont{Lin}}, \bibinfo{author}{\bibfnamefont{I.}~\bibnamefont{Teper}}, \bibinfo{author}{\bibfnamefont{C.}~\bibnamefont{Chin}}, \bibnamefont{and} \bibinfo{author}{\bibfnamefont{V.}~\bibnamefont{Vuleti{\'c}}}, \bibinfo{journal}{Physical Review Letters} \textbf{\bibinfo{volume}{92}}, \bibinfo{pages}{050404} (\bibinfo{year}{2004}).

\bibitem[{\citenamefont{Braaten et~al.}(2002)\citenamefont{Braaten, Hammer, and Mehen}}]{PhysRevLett.88.040401}
\bibinfo{author}{\bibfnamefont{E.}~\bibnamefont{Braaten}}, \bibinfo{author}{\bibfnamefont{H.-W.} \bibnamefont{Hammer}}, \bibnamefont{and} \bibinfo{author}{\bibfnamefont{T.}~\bibnamefont{Mehen}}, \bibinfo{journal}{Phys. Rev. Lett.} \textbf{\bibinfo{volume}{88}}, \bibinfo{pages}{040401} (\bibinfo{year}{2002}), \urlprefix\url{https://link.aps.org/doi/10.1103/PhysRevLett.88.040401}.

\bibitem[{\citenamefont{Tomza et~al.}(2019)\citenamefont{Tomza, Jachymski, Gerritsma, Negretti, Calarco, Idziaszek, and Julienne}}]{tomza2019cold}
\bibinfo{author}{\bibfnamefont{M.}~\bibnamefont{Tomza}}, \bibinfo{author}{\bibfnamefont{K.}~\bibnamefont{Jachymski}}, \bibinfo{author}{\bibfnamefont{R.}~\bibnamefont{Gerritsma}}, \bibinfo{author}{\bibfnamefont{A.}~\bibnamefont{Negretti}}, \bibinfo{author}{\bibfnamefont{T.}~\bibnamefont{Calarco}}, \bibinfo{author}{\bibfnamefont{Z.}~\bibnamefont{Idziaszek}}, \bibnamefont{and} \bibinfo{author}{\bibfnamefont{P.~S.} \bibnamefont{Julienne}}, \bibinfo{journal}{Reviews of modern physics} \textbf{\bibinfo{volume}{91}}, \bibinfo{pages}{035001} (\bibinfo{year}{2019}).

\bibitem[{\citenamefont{{\'S}mia{\l}kowski and Tomza}(2020)}]{smialkowski2020}
\bibinfo{author}{\bibfnamefont{M.}~\bibnamefont{{\'S}mia{\l}kowski}} \bibnamefont{and} \bibinfo{author}{\bibfnamefont{M.}~\bibnamefont{Tomza}}, \bibinfo{journal}{Physical Review A} \textbf{\bibinfo{volume}{101}}, \bibinfo{pages}{012501} (\bibinfo{year}{2020}).

\bibitem[{\citenamefont{Wang and Ye}(2007)}]{wang2007}
\bibinfo{author}{\bibfnamefont{G.}~\bibnamefont{Wang}} \bibnamefont{and} \bibinfo{author}{\bibfnamefont{A.}~\bibnamefont{Ye}}, \bibinfo{journal}{Physical Review A} \textbf{\bibinfo{volume}{76}}, \bibinfo{pages}{043409} (\bibinfo{year}{2007}).

\bibitem[{\citenamefont{Brickman et~al.}(2007)\citenamefont{Brickman, Chang, Acton, Chew, Matsukevich, Haljan, Bagnato, and Monroe}}]{brickman2007}
\bibinfo{author}{\bibfnamefont{K.-A.} \bibnamefont{Brickman}}, \bibinfo{author}{\bibfnamefont{M.-S.} \bibnamefont{Chang}}, \bibinfo{author}{\bibfnamefont{M.}~\bibnamefont{Acton}}, \bibinfo{author}{\bibfnamefont{A.}~\bibnamefont{Chew}}, \bibinfo{author}{\bibfnamefont{D.}~\bibnamefont{Matsukevich}}, \bibinfo{author}{\bibfnamefont{P.}~\bibnamefont{Haljan}}, \bibinfo{author}{\bibfnamefont{V.~S.} \bibnamefont{Bagnato}}, \bibnamefont{and} \bibinfo{author}{\bibfnamefont{C.}~\bibnamefont{Monroe}}, \bibinfo{journal}{Physical Review A} \textbf{\bibinfo{volume}{76}}, \bibinfo{pages}{043411} (\bibinfo{year}{2007}).

\bibitem[{\citenamefont{Hachisu et~al.}(2008)\citenamefont{Hachisu, Miyagishi, Porsev, Derevianko, Ovsiannikov, Pal’chikov, Takamoto, and Katori}}]{hachisu2008trapping}
\bibinfo{author}{\bibfnamefont{H.}~\bibnamefont{Hachisu}}, \bibinfo{author}{\bibfnamefont{K.}~\bibnamefont{Miyagishi}}, \bibinfo{author}{\bibfnamefont{S.}~\bibnamefont{Porsev}}, \bibinfo{author}{\bibfnamefont{A.}~\bibnamefont{Derevianko}}, \bibinfo{author}{\bibfnamefont{V.}~\bibnamefont{Ovsiannikov}}, \bibinfo{author}{\bibfnamefont{.~f.~V.} \bibnamefont{Pal’chikov}}, \bibinfo{author}{\bibfnamefont{M.}~\bibnamefont{Takamoto}}, \bibnamefont{and} \bibinfo{author}{\bibfnamefont{H.}~\bibnamefont{Katori}}, \bibinfo{journal}{Physical Review Letters} \textbf{\bibinfo{volume}{100}}, \bibinfo{pages}{053001} (\bibinfo{year}{2008}).

\bibitem[{\citenamefont{Yan et~al.}(2020)\citenamefont{Yan, Tang, Yan, and Babb}}]{PhysRevA.101.032702}
\bibinfo{author}{\bibfnamefont{P.-G.} \bibnamefont{Yan}}, \bibinfo{author}{\bibfnamefont{L.-Y.} \bibnamefont{Tang}}, \bibinfo{author}{\bibfnamefont{Z.-C.} \bibnamefont{Yan}}, \bibnamefont{and} \bibinfo{author}{\bibfnamefont{J.~F.} \bibnamefont{Babb}}, \bibinfo{journal}{Phys. Rev. A} \textbf{\bibinfo{volume}{101}}, \bibinfo{pages}{032702} (\bibinfo{year}{2020}), \urlprefix\url{https://link.aps.org/doi/10.1103/PhysRevA.101.032702}.

\bibitem[{\citenamefont{Kar and Ho}(2010)}]{PhysRevA.81.062506}
\bibinfo{author}{\bibfnamefont{S.}~\bibnamefont{Kar}} \bibnamefont{and} \bibinfo{author}{\bibfnamefont{Y.~K.} \bibnamefont{Ho}}, \bibinfo{journal}{Phys. Rev. A} \textbf{\bibinfo{volume}{81}}, \bibinfo{pages}{062506} (\bibinfo{year}{2010}), \urlprefix\url{https://link.aps.org/doi/10.1103/PhysRevA.81.062506}.

\bibitem[{\citenamefont{Kar et~al.}(2013)\citenamefont{Kar, Li, and Shen}}]{KAR201334}
\bibinfo{author}{\bibfnamefont{S.}~\bibnamefont{Kar}}, \bibinfo{author}{\bibfnamefont{H.}~\bibnamefont{Li}}, \bibnamefont{and} \bibinfo{author}{\bibfnamefont{Z.}~\bibnamefont{Shen}}, \bibinfo{journal}{Journal of Quantitative Spectroscopy and Radiative Transfer} \textbf{\bibinfo{volume}{116}}, \bibinfo{pages}{34} (\bibinfo{year}{2013}), ISSN \bibinfo{issn}{0022-4073}, \urlprefix\url{https://www.sciencedirect.com/science/article/pii/S0022407312004918}.

\bibitem[{\citenamefont{Lu et~al.}(2025)\citenamefont{Lu, Zheng, Yan, Babb, and Tang}}]{Shan-Shan.Lu.23202}
\bibinfo{author}{\bibfnamefont{S.-S.} \bibnamefont{Lu}}, \bibinfo{author}{\bibfnamefont{H.-Y.} \bibnamefont{Zheng}}, \bibinfo{author}{\bibfnamefont{Z.-C.} \bibnamefont{Yan}}, \bibinfo{author}{\bibfnamefont{J.~F.} \bibnamefont{Babb}}, \bibnamefont{and} \bibinfo{author}{\bibfnamefont{L.-Y.} \bibnamefont{Tang}}, \bibinfo{journal}{Chinese Physics B} \textbf{\bibinfo{volume}{34}}, \bibinfo{eid}{023202} (\bibinfo{year}{2025}), \urlprefix\url{https://cpb.iphy.ac.cn/EN/abstract/article_127362.shtml}.

\bibitem[{\citenamefont{Gr\"unert and Hemmerich}(2002)}]{PhysRevA.65.041401}
\bibinfo{author}{\bibfnamefont{J.}~\bibnamefont{Gr\"unert}} \bibnamefont{and} \bibinfo{author}{\bibfnamefont{A.}~\bibnamefont{Hemmerich}}, \bibinfo{journal}{Phys. Rev. A} \textbf{\bibinfo{volume}{65}}, \bibinfo{pages}{041401} (\bibinfo{year}{2002}), \urlprefix\url{https://link.aps.org/doi/10.1103/PhysRevA.65.041401}.

\bibitem[{\citenamefont{Tang et~al.}(2012)\citenamefont{Tang, Yan, Shi, Babb, and Mitroy}}]{doi:10.1063/1.3691891}
\bibinfo{author}{\bibfnamefont{L.-Y.} \bibnamefont{Tang}}, \bibinfo{author}{\bibfnamefont{Z.-C.} \bibnamefont{Yan}}, \bibinfo{author}{\bibfnamefont{T.-Y.} \bibnamefont{Shi}}, \bibinfo{author}{\bibfnamefont{J.~F.} \bibnamefont{Babb}}, \bibnamefont{and} \bibinfo{author}{\bibfnamefont{J.}~\bibnamefont{Mitroy}}, \bibinfo{journal}{The Journal of Chemical Physics} \textbf{\bibinfo{volume}{136}}, \bibinfo{pages}{104104} (\bibinfo{year}{2012}).

\bibitem[{\citenamefont{Derevianko et~al.}(2001)\citenamefont{Derevianko, Babb, and Dalgarno}}]{derevianko2001high}
\bibinfo{author}{\bibfnamefont{A.}~\bibnamefont{Derevianko}}, \bibinfo{author}{\bibfnamefont{J.}~\bibnamefont{Babb}}, \bibnamefont{and} \bibinfo{author}{\bibfnamefont{A.}~\bibnamefont{Dalgarno}}, \bibinfo{journal}{Physical Review A} \textbf{\bibinfo{volume}{63}}, \bibinfo{pages}{052704} (\bibinfo{year}{2001}).

\bibitem[{\citenamefont{Porsev and Derevianko}(2002)}]{porsev2002high}
\bibinfo{author}{\bibfnamefont{S.~G.} \bibnamefont{Porsev}} \bibnamefont{and} \bibinfo{author}{\bibfnamefont{A.}~\bibnamefont{Derevianko}}, \bibinfo{journal}{Physical Review A} \textbf{\bibinfo{volume}{65}}, \bibinfo{pages}{020701} (\bibinfo{year}{2002}).

\bibitem[{\citenamefont{Lima and Caldas}(2005)}]{PhysRevB.72.033109}
\bibinfo{author}{\bibfnamefont{N.~A.} \bibnamefont{Lima}} \bibnamefont{and} \bibinfo{author}{\bibfnamefont{M.~J.} \bibnamefont{Caldas}}, \bibinfo{journal}{Phys. Rev. B} \textbf{\bibinfo{volume}{72}}, \bibinfo{pages}{033109} (\bibinfo{year}{2005}), \urlprefix\url{https://link.aps.org/doi/10.1103/PhysRevB.72.033109}.

\bibitem[{\citenamefont{Shukla et~al.}(2022)\citenamefont{Shukla, Kaur, Arora, and Srivastava}}]{shukla2022two}
\bibinfo{author}{\bibfnamefont{N.}~\bibnamefont{Shukla}}, \bibinfo{author}{\bibfnamefont{H.}~\bibnamefont{Kaur}}, \bibinfo{author}{\bibfnamefont{B.}~\bibnamefont{Arora}}, \bibnamefont{and} \bibinfo{author}{\bibfnamefont{R.}~\bibnamefont{Srivastava}}, \bibinfo{journal}{Physica B: Condensed Matter} \textbf{\bibinfo{volume}{624}}, \bibinfo{pages}{413422} (\bibinfo{year}{2022}).

\bibitem[{\citenamefont{Brahms et~al.}(2008)\citenamefont{Brahms, Newman, Johnson, Greytak, Kleppner, and Doyle}}]{PhysRevLett.101.103002}
\bibinfo{author}{\bibfnamefont{N.}~\bibnamefont{Brahms}}, \bibinfo{author}{\bibfnamefont{B.}~\bibnamefont{Newman}}, \bibinfo{author}{\bibfnamefont{C.}~\bibnamefont{Johnson}}, \bibinfo{author}{\bibfnamefont{T.}~\bibnamefont{Greytak}}, \bibinfo{author}{\bibfnamefont{D.}~\bibnamefont{Kleppner}}, \bibnamefont{and} \bibinfo{author}{\bibfnamefont{J.}~\bibnamefont{Doyle}}, \bibinfo{journal}{Phys. Rev. Lett.} \textbf{\bibinfo{volume}{101}}, \bibinfo{pages}{103002} (\bibinfo{year}{2008}), \urlprefix\url{https://link.aps.org/doi/10.1103/PhysRevLett.101.103002}.

\bibitem[{\citenamefont{Uhlenberg et~al.}(2000)\citenamefont{Uhlenberg, Dirscherl, and Walther}}]{PhysRevA.62.063404}
\bibinfo{author}{\bibfnamefont{G.}~\bibnamefont{Uhlenberg}}, \bibinfo{author}{\bibfnamefont{J.}~\bibnamefont{Dirscherl}}, \bibnamefont{and} \bibinfo{author}{\bibfnamefont{H.}~\bibnamefont{Walther}}, \bibinfo{journal}{Phys. Rev. A} \textbf{\bibinfo{volume}{62}}, \bibinfo{pages}{063404} (\bibinfo{year}{2000}), \urlprefix\url{https://link.aps.org/doi/10.1103/PhysRevA.62.063404}.

\bibitem[{hen(1999)}]{henkel1999loss}
\bibinfo{journal}{Applied Physics B} \textbf{\bibinfo{volume}{69}}, \bibinfo{pages}{379} (\bibinfo{year}{1999}).

\bibitem[{\citenamefont{McEnery et~al.}(2013)\citenamefont{McEnery, {\"O}zdemir, and Kim}}]{tame2013quantum}
\bibinfo{author}{\bibfnamefont{K.}~\bibnamefont{McEnery}}, \bibinfo{author}{\bibfnamefont{{\c{S}}.}~\bibnamefont{{\"O}zdemir}}, \bibnamefont{and} \bibinfo{author}{\bibfnamefont{M.}~\bibnamefont{Kim}}, \bibinfo{journal}{Nature Physics} \textbf{\bibinfo{volume}{9}}, \bibinfo{pages}{329} (\bibinfo{year}{2013}).

\bibitem[{\citenamefont{Reilly and Tkatchenko}(2015)}]{reilly2015van}
\bibinfo{author}{\bibfnamefont{A.~M.} \bibnamefont{Reilly}} \bibnamefont{and} \bibinfo{author}{\bibfnamefont{A.}~\bibnamefont{Tkatchenko}}, \bibinfo{journal}{Chemical Science} \textbf{\bibinfo{volume}{6}}, \bibinfo{pages}{3289} (\bibinfo{year}{2015}).

\bibitem[{\citenamefont{Mitroy and Zhang}(2008)}]{10.1063/1.2841470}
\bibinfo{author}{\bibfnamefont{J.}~\bibnamefont{Mitroy}} \bibnamefont{and} \bibinfo{author}{\bibfnamefont{J.-Y.} \bibnamefont{Zhang}}, \bibinfo{journal}{The Journal of Chemical Physics} \textbf{\bibinfo{volume}{128}}, \bibinfo{pages}{134305} (\bibinfo{year}{2008}), ISSN \bibinfo{issn}{0021-9606}, \eprint{https://pubs.aip.org/aip/jcp/article-pdf/doi/10.1063/1.2841470/13876386/134305\_1\_online.pdf}, \urlprefix\url{https://doi.org/10.1063/1.2841470}.

\bibitem[{\citenamefont{Arora et~al.}(2012)\citenamefont{Arora, Nandy, and Sahoo}}]{arora2012multipolar}
\bibinfo{author}{\bibfnamefont{B.}~\bibnamefont{Arora}}, \bibinfo{author}{\bibfnamefont{D.}~\bibnamefont{Nandy}}, \bibnamefont{and} \bibinfo{author}{\bibfnamefont{B.}~\bibnamefont{Sahoo}}, \bibinfo{journal}{Physical Review A} \textbf{\bibinfo{volume}{85}}, \bibinfo{pages}{012506} (\bibinfo{year}{2012}).

\bibitem[{\citenamefont{Shukla et~al.}(2020)\citenamefont{Shukla, Arora, Sharma, and Srivastava}}]{shukla2020two}
\bibinfo{author}{\bibfnamefont{N.}~\bibnamefont{Shukla}}, \bibinfo{author}{\bibfnamefont{B.}~\bibnamefont{Arora}}, \bibinfo{author}{\bibfnamefont{L.}~\bibnamefont{Sharma}}, \bibnamefont{and} \bibinfo{author}{\bibfnamefont{R.}~\bibnamefont{Srivastava}}, \bibinfo{journal}{Physical Review A} \textbf{\bibinfo{volume}{102}}, \bibinfo{pages}{022817} (\bibinfo{year}{2020}).

\bibitem[{\citenamefont{Arora et~al.}(2007)\citenamefont{Arora, Safronova, and Clark}}]{arora2007magic}
\bibinfo{author}{\bibfnamefont{B.}~\bibnamefont{Arora}}, \bibinfo{author}{\bibfnamefont{M.}~\bibnamefont{Safronova}}, \bibnamefont{and} \bibinfo{author}{\bibfnamefont{C.~W.} \bibnamefont{Clark}}, \bibinfo{journal}{Physical Review A—Atomic, Molecular, and Optical Physics} \textbf{\bibinfo{volume}{76}}, \bibinfo{pages}{052509} (\bibinfo{year}{2007}).

\bibitem[{\citenamefont{Safronova et~al.}(1999)\citenamefont{Safronova, Johnson, and Derevianko}}]{PhysRevA.60.4476}
\bibinfo{author}{\bibfnamefont{M.~S.} \bibnamefont{Safronova}}, \bibinfo{author}{\bibfnamefont{W.~R.} \bibnamefont{Johnson}}, \bibnamefont{and} \bibinfo{author}{\bibfnamefont{A.}~\bibnamefont{Derevianko}}, \bibinfo{journal}{Phys. Rev. A} \textbf{\bibinfo{volume}{60}}, \bibinfo{pages}{4476} (\bibinfo{year}{1999}), \urlprefix\url{https://link.aps.org/doi/10.1103/PhysRevA.60.4476}.

\bibitem[{\citenamefont{Johnson}(2007)}]{johnson2007atomic}
\bibinfo{author}{\bibfnamefont{W.~R.} \bibnamefont{Johnson}}, \emph{\bibinfo{title}{Atomic structure theory}} (\bibinfo{publisher}{Springer}, \bibinfo{year}{2007}).

\bibitem[{\citenamefont{Blundell et~al.}(1987)\citenamefont{Blundell, Guo, Johnson, and Sapirstein}}]{blundell1987formulas}
\bibinfo{author}{\bibfnamefont{S.}~\bibnamefont{Blundell}}, \bibinfo{author}{\bibfnamefont{D.}~\bibnamefont{Guo}}, \bibinfo{author}{\bibfnamefont{W.}~\bibnamefont{Johnson}}, \bibnamefont{and} \bibinfo{author}{\bibfnamefont{J.}~\bibnamefont{Sapirstein}}, \bibinfo{journal}{Atomic data and nuclear data tables} \textbf{\bibinfo{volume}{37}}, \bibinfo{pages}{103} (\bibinfo{year}{1987}).

\bibitem[{\citenamefont{Johnson et~al.}(1996)\citenamefont{Johnson, Liu, and Sapirstein}}]{johnson1996transition}
\bibinfo{author}{\bibfnamefont{W.}~\bibnamefont{Johnson}}, \bibinfo{author}{\bibfnamefont{Z.}~\bibnamefont{Liu}}, \bibnamefont{and} \bibinfo{author}{\bibfnamefont{J.}~\bibnamefont{Sapirstein}}, \bibinfo{journal}{Atomic Data and Nuclear Data Tables} \textbf{\bibinfo{volume}{64}}, \bibinfo{pages}{279} (\bibinfo{year}{1996}).

\bibitem[{\citenamefont{Safronova et~al.}(2005)\citenamefont{Safronova, Safronova, and Johnson}}]{safronova2005excitation}
\bibinfo{author}{\bibfnamefont{U.~I.} \bibnamefont{Safronova}}, \bibinfo{author}{\bibfnamefont{M.~S.} \bibnamefont{Safronova}}, \bibnamefont{and} \bibinfo{author}{\bibfnamefont{W.~R.} \bibnamefont{Johnson}}, \bibinfo{journal}{Phys. Rev. A} \textbf{\bibinfo{volume}{71}}, \bibinfo{pages}{052506} (\bibinfo{year}{2005}), \urlprefix\url{https://link.aps.org/doi/10.1103/PhysRevA.71.052506}.

\bibitem[{\citenamefont{Lindgren and Morrison}(2012)}]{lindgren2012atomic}
\bibinfo{author}{\bibfnamefont{I.}~\bibnamefont{Lindgren}} \bibnamefont{and} \bibinfo{author}{\bibfnamefont{J.}~\bibnamefont{Morrison}}, \emph{\bibinfo{title}{Atomic many-body theory}}, vol.~\bibinfo{volume}{3} (\bibinfo{publisher}{Springer Science \& Business Media}, \bibinfo{year}{2012}).

\bibitem[{\citenamefont{Neogrády et~al.}(1997)\citenamefont{Neogrády, Kellö, Urban, and Sadlej}}]{ijqc}
\bibinfo{author}{\bibfnamefont{P.}~\bibnamefont{Neogrády}}, \bibinfo{author}{\bibfnamefont{V.}~\bibnamefont{Kellö}}, \bibinfo{author}{\bibfnamefont{M.}~\bibnamefont{Urban}}, \bibnamefont{and} \bibinfo{author}{\bibfnamefont{A.~J.} \bibnamefont{Sadlej}}, \bibinfo{journal}{International Journal of Quantum Chemistry} \textbf{\bibinfo{volume}{63}}, \bibinfo{pages}{557} (\bibinfo{year}{1997}).

\bibitem[{\citenamefont{Zhang et~al.}(2008)\citenamefont{Zhang, Mitroy, Sadeghpour, and Bromley}}]{PhysRevA.78.062710}
\bibinfo{author}{\bibfnamefont{J.~Y.} \bibnamefont{Zhang}}, \bibinfo{author}{\bibfnamefont{J.}~\bibnamefont{Mitroy}}, \bibinfo{author}{\bibfnamefont{H.~R.} \bibnamefont{Sadeghpour}}, \bibnamefont{and} \bibinfo{author}{\bibfnamefont{M.~W.~J.} \bibnamefont{Bromley}}, \bibinfo{journal}{Phys. Rev. A} \textbf{\bibinfo{volume}{78}}, \bibinfo{pages}{062710} (\bibinfo{year}{2008}), \urlprefix\url{https://link.aps.org/doi/10.1103/PhysRevA.78.062710}.

\bibitem[{\citenamefont{Gould and Bucko}(2016)}]{gould2016c}
\bibinfo{author}{\bibfnamefont{T.}~\bibnamefont{Gould}} \bibnamefont{and} \bibinfo{author}{\bibfnamefont{T.}~\bibnamefont{Bucko}}, \bibinfo{journal}{Journal of chemical theory and computation} \textbf{\bibinfo{volume}{12}}, \bibinfo{pages}{3603} (\bibinfo{year}{2016}).

\bibitem[{\citenamefont{{\'S}mia{\l}kowski and Tomza}(2021)}]{smialkowski2021highly}
\bibinfo{author}{\bibfnamefont{M.}~\bibnamefont{{\'S}mia{\l}kowski}} \bibnamefont{and} \bibinfo{author}{\bibfnamefont{M.}~\bibnamefont{Tomza}}, \bibinfo{journal}{Physical Review A} \textbf{\bibinfo{volume}{103}}, \bibinfo{pages}{022802} (\bibinfo{year}{2021}).

\bibitem[{\citenamefont{Sarkisov}(2022)}]{10.1063/5.0084981}
\bibinfo{author}{\bibfnamefont{G.~S.} \bibnamefont{Sarkisov}}, \bibinfo{journal}{Physics of Plasmas} \textbf{\bibinfo{volume}{29}}, \bibinfo{pages}{073502} (\bibinfo{year}{2022}), ISSN \bibinfo{issn}{1070-664X}, \eprint{https://pubs.aip.org/aip/pop/article-pdf/doi/10.1063/5.0084981/16592973/073502\_1\_online.pdf}, \urlprefix\url{https://doi.org/10.1063/5.0084981}.

\bibitem[{\citenamefont{J{\"o}nsson et~al.}(2013)\citenamefont{J{\"o}nsson, Gaigalas, Biero{\'n}, Fischer, and Grant}}]{jonsson2013new}
\bibinfo{author}{\bibfnamefont{P.}~\bibnamefont{J{\"o}nsson}}, \bibinfo{author}{\bibfnamefont{G.}~\bibnamefont{Gaigalas}}, \bibinfo{author}{\bibfnamefont{J.}~\bibnamefont{Biero{\'n}}}, \bibinfo{author}{\bibfnamefont{C.~F.} \bibnamefont{Fischer}}, \bibnamefont{and} \bibinfo{author}{\bibfnamefont{I.}~\bibnamefont{Grant}}, \bibinfo{journal}{Computer Physics Communications} \textbf{\bibinfo{volume}{184}}, \bibinfo{pages}{2197} (\bibinfo{year}{2013}).

\bibitem[{\citenamefont{Barakhshan et~al.}()\citenamefont{Barakhshan, Marrs, Bhosale, Arora, Eigenmann, and Safronova}}]{UDportal}
\bibinfo{author}{\bibfnamefont{P.}~\bibnamefont{Barakhshan}}, \bibinfo{author}{\bibfnamefont{A.}~\bibnamefont{Marrs}}, \bibinfo{author}{\bibfnamefont{A.}~\bibnamefont{Bhosale}}, \bibinfo{author}{\bibfnamefont{B.}~\bibnamefont{Arora}}, \bibinfo{author}{\bibfnamefont{R.}~\bibnamefont{Eigenmann}}, \bibnamefont{and} \bibinfo{author}{\bibfnamefont{M.~S.} \bibnamefont{Safronova}}, \bibinfo{howpublished}{Portal for High-Precision Atomic Data and Computation (version 2.0). University of Delaware, Newark, DE, USA}, \urlprefix\url{https://www.udel.edu/atom [February 2022].}

\bibitem[{\citenamefont{Kaur et~al.}(2021)\citenamefont{Kaur, Shukla, Srivastava, and Arora}}]{kaur2021dispersion}
\bibinfo{author}{\bibfnamefont{H.}~\bibnamefont{Kaur}}, \bibinfo{author}{\bibfnamefont{N.}~\bibnamefont{Shukla}}, \bibinfo{author}{\bibfnamefont{R.}~\bibnamefont{Srivastava}}, \bibnamefont{and} \bibinfo{author}{\bibfnamefont{B.}~\bibnamefont{Arora}}, \bibinfo{journal}{Physical Review A} \textbf{\bibinfo{volume}{104}}, \bibinfo{pages}{012806} (\bibinfo{year}{2021}).

\bibitem[{\citenamefont{Kramida et~al.}(2020)\citenamefont{Kramida, {Yu.~Ralchenko}, Reader, and {and NIST ASD Team}}}]{NIST_ASD}
\bibinfo{author}{\bibfnamefont{A.}~\bibnamefont{Kramida}}, \bibinfo{author}{\bibnamefont{{Yu.~Ralchenko}}}, \bibinfo{author}{\bibfnamefont{J.}~\bibnamefont{Reader}}, \bibnamefont{and} \bibinfo{author}{\bibnamefont{{and NIST ASD Team}}}, \bibinfo{howpublished}{{NIST Atomic Spectra Database (ver. 5.8), [Online]. Available: {\tt{https://physics.nist.gov/asd}} [2021, July 17]. National Institute of Standards and Technology, Gaithersburg, MD.}} (\bibinfo{year}{2020}).

\bibitem[{\citenamefont{Burke}(2012)}]{burke2012perspective}
\bibinfo{author}{\bibfnamefont{K.}~\bibnamefont{Burke}}, \bibinfo{journal}{The Journal of chemical physics} \textbf{\bibinfo{volume}{136}} (\bibinfo{year}{2012}).

\bibitem[{\citenamefont{Ma et~al.}(2015)\citenamefont{Ma, Indergaard, Zhang, Larkin, Moro, and de~Heer}}]{PhysRevA.91.010501}
\bibinfo{author}{\bibfnamefont{L.}~\bibnamefont{Ma}}, \bibinfo{author}{\bibfnamefont{J.}~\bibnamefont{Indergaard}}, \bibinfo{author}{\bibfnamefont{B.}~\bibnamefont{Zhang}}, \bibinfo{author}{\bibfnamefont{I.}~\bibnamefont{Larkin}}, \bibinfo{author}{\bibfnamefont{R.}~\bibnamefont{Moro}}, \bibnamefont{and} \bibinfo{author}{\bibfnamefont{W.~A.} \bibnamefont{de~Heer}}, \bibinfo{journal}{Phys. Rev. A} \textbf{\bibinfo{volume}{91}}, \bibinfo{pages}{010501} (\bibinfo{year}{2015}), \urlprefix\url{https://link.aps.org/doi/10.1103/PhysRevA.91.010501}.

\end{thebibliography}

\end{document}